\pdfoutput=1

\documentclass[11pt,twoside,a4paper,cmspaper,final,collab]{cms-tdr}
\def\svnVersion{46728:46809M}\def\cmsCernNo{2011-020}\def\cmsCernDate{\today}\def\cmsMessage{Submitted to the Journal of High Energy Physics}

\begin{document}\cmsNoteHeader{EXO-10-026}

\hyphenation{had-ron-i-za-tion}
\hyphenation{cal-or-i-me-ter}
\hyphenation{de-vices}
\RCS$Revision: 46809 $
\RCS$HeadURL: svn+ssh://alverson@svn.cern.ch/reps/tdr2/papers/EXO-10-026/trunk/EXO-10-026.tex $
\RCS$Id: EXO-10-026.tex 46809 2011-03-22 11:47:23Z alverson $
\cmsNoteHeader{EXO-10-026} 
\title{Search for Large Extra Dimensions in the Diphoton Final State at the Large Hadron Collider}

\address[Brown]{Brown University, Department of Physics, 182 Hope St., Providence, RI 02912, USA}
\author[Brown]{John Paul Chou}
\author[Brown]{Greg Landsberg}
\author[Brown]{Duong Nguyen}

\date{\today}

\abstract{
A search for large extra spatial dimensions via virtual-graviton exchange in the diphoton channel has been carried out with the CMS detector at the LHC. No excess of events above the standard model expectations is found using a data sample collected in proton-proton collisions at $\sqrt{s} = 7$~TeV and corresponding to an integrated luminosity of 36 pb$^{-1}$.  New lower limits on the effective Planck scale in the range of 1.6--2.3 TeV at the 95\% confidence level are set, providing the most restrictive bounds to date on models with more than two large extra dimensions.}

\hypersetup{%
pdfauthor={CMS Collaboration},%
pdftitle={Search for Large Extra Dimensions in the Diphoton Final State at the Large Hadron Collider},%
pdfsubject={CMS},%
pdfkeywords={CMS, physics, exotica}}

\maketitle 

\section{Introduction}

Compact large extra dimensions (ED) are an intriguing proposed solution to the hierarchy problem of the standard model (SM), which refers to the puzzling fact that the fundamental scale of gravity $M_{\rm Pl}\sim10^{19}$~GeV is so much higher than the electroweak symmetry breaking scale $\sim10^3$~GeV.  With such a difference in scales, it is difficult to protect the Higgs boson mass from radiative corrections without a very high degree of fine-tuning. The original proposal to use ED to solve the hierarchy problem was presented by Arkani-Hamed, Dimopoulos, and Dvali (ADD)~\cite{add,add1}.  They posited a scenario wherein the SM is constrained to the common 3+1 space-time dimensions, while gravity is free to propagate through the entire multidimensional space. Thus, the gravitational flux in 3+1 dimensions is effectively diluted by virtue of the multidimensional Gauss's Law.  The fundamental Planck scale $M_{\rm D}$ is therefore related to the apparent scale $M_{\rm Pl}$ according to the formula $M_{\rm D}^{n_{\rm ED}+2}=M_{\rm Pl}^2/r^{n_{\rm ED}}$, where $r$ and $n_{\rm ED}$ are the size and number of the extra dimensions, respectively. A review of current limits on the ADD model can be found elsewhere~\cite{PDG}.

Phenomenologically, this large ED scenario results in $s$-channel production of massive, virtual Kaluza--Klein (KK) graviton states, which can decay into two photons~\cite{KCGL}.  Decays to fermions are suppressed relative to photons because the graviton is spin-2 and the fermions cannot be produced in the $s$ wave.  Because the KK gravitons propagate through the compact extra dimensions, their wavefunction must satisfy periodic boundary conditions, which result in discrete energy levels with modal spacing on the order of the inverse ED size, from 1~meV to 100~MeV.  This results in an apparent continuum spectrum of diphotons, rather than distinct resonances.

Summing over KK modes results in a divergence in the cross section, so an ultraviolet (UV) cutoff scale $M_{\rm S}$ is needed.  This scale is related to---but potentially different from---the fundamental Planck scale $M_{\rm D}$, as the precise relationship depends on the UV completion of the effective theory.  The effects of virtual-graviton production on the cross section are parameterized by the single variable $\eta_G=\mathcal{F}/M_{\rm S}^4$, where $\mathcal{F}$ is an order-unity dimensionless parameter, for which several conventional assumptions exist:
\begin{eqnarray}
\mathcal{F}&=&1\label{eq:GRW}~~~\mbox{(Guidice, Rattazzi, and Wells, GRW~\cite{GRW}),} \\
\mathcal{F}& =& \left\{ \begin{array}{cr}
\log\left(\frac{M_{\rm S}^2}{\hat{s}}\right) & \mbox{if $n_{\rm ED}=2$} \smallskip\\
\frac{2}{(n_{\rm ED}-2)} & \mbox{if $n_{\rm ED}>2$}\end{array}\right.\label{eq:HLZ}~~~\mbox{(Han, Lykken, and Zhang, HLZ \cite{HLZ}),} \\
\mathcal{F}&=&\pm\frac{2}{\pi}\label{eq:Hewett}~~~\mbox{(Hewett \cite{Hewett}),}
\end{eqnarray}
where $\sqrt{\hat{s}}$ is the center-of-mass energy of the hard parton-parton collision.  We note that the HLZ convention contains an explicit dependence on the number of extra dimensions.

Searches for extra dimensions via virtual-graviton effects have been conducted at HERA, LEP, and the Tevatron (Refs.~\cite{PDG,GL} contain recent reviews of these searches). The most stringent previously published limits on $M_{\rm S}$ come from the D0 measurements in the dijet \cite{D0dijet} and diphoton plus dielectron \cite{D0diphoton} channels, which exclude values of $M_{\rm S}$ lower than 1.3$-$2.1 TeV at 95\% confidence level (CL), depending on $n_{\rm ED}$.

In this paper, we present a search for virtual-graviton contributions in the diphoton final state, using a data sample corresponding to an integrated luminosity of $36$~pb$^{-1}$, collected in pp collisions at $\sqrt{s} = 7$ TeV at the CERN Large Hadron Collider (LHC) with the Compact Muon Solenoid (CMS) detector.

\section{The CMS Detector}

CMS is a general-purpose detector designed to study proton collisions at the LHC and is described in detail in Ref.~\cite{CMS}. The detector consists of an all-silicon tracker, an electromagnetic calorimeter (ECAL), and a hadronic sampling calorimeter (HCAL), all contained inside a large-bore 3.8 T superconducting solenoid.  In the central region, the tracker consists of three layers of silicon pixel detectors, followed by 10 layers of single- and double-sided silicon-strip detectors. The calorimeter towers are projective and finely segmented, with $\Delta\eta\approx\Delta\phi\approx 0.087$ in the central region. Moreover, each tower consists of a five-by-five transverse grid of ECAL crystals ($\Delta\eta\approx\Delta\phi\approx 0.0174$), allowing precise reconstruction of the e/$\gamma$ position and energy. Here, the pseudorapidity $\eta$ is defined as $-\ln(\tan\frac{\theta}{2})$, where $\theta$ is the polar angle with respect to the direction of the counterclockwise beam, and $\phi$ is the azimuthal angle. Beyond the solenoid there are four layers of muon detectors, which are interspersed throughout the steel return yoke of the magnet.  The instantaneous luminosity is measured with a relative uncertainty of 4\% using information from forward hadronic calorimeters~\cite{lumi}.

The CMS trigger system consists of two levels. The first level (L1), composed of custom hardware, uses information from the calorimeters and muon detectors to select the most interesting events for more refined selection and analysis at a rate of up to 80~kHz. The software-based High Level Trigger further decreases the rate to a maximum of $\sim 300$~Hz for data storage.  Events for the control samples used in this analysis were collected through a single-photon trigger, where the photon was required to have a transverse energy $\et\equiv E\sin\theta$ of at least 20 or 30 GeV, depending on the data collection period.  Events in the signal sample were also initially collected through a single $\et>20$~GeV photon trigger.  However, as the instantaneous luminosity increased, we switched to a double-photon trigger requiring both photons to have $\et>22$~GeV.

\section{Event Reconstruction and Selection}

We first require that an event be consistent with a pp collision and have at least one well-reconstructed primary vertex~\cite{CMSVertex}.  We then reconstruct photons with $\et>30$~GeV in the ECAL barrel fiducial region ($|\eta|<1.44$) by clustering electromagnetic energy depositions in the ECAL. The ECAL clusters are five crystals wide in $\eta$ and a variable length in $\phi$ to capture associated electromagnetic energy from possible photon conversions in the tracker. If hits are present in the pixel detector consistent with an electron track whose momentum and location is similar to the energy and location of the ECAL cluster, then the cluster is rejected as a photon candidate.  In 2010 collision data, the ECAL has an energy resolution better than 1\% in the barrel for unconverted photons with $\et>20$~GeV~\cite{ECALPAS}.

Hadronic jets can be misidentified as photons when their leading hadron is a hard $\pi^0$ or $\eta$.  We reduce the misidentification rate by placing the following restrictions on the isolation of the cluster: (i) the hadronic energy within $\Delta R<0.15$ of the cluster must be less than $5\%$ of its electromagnetic energy; (ii) the scalar sum of the transverse momentum of tracks, $\Sigma \pt$, associated with the primary event vertex surrounding the cluster within a hollow cone of $0.04<\Delta R<0.40$ must be less than $2.0\GeV + 0.001\et$, where $\et$ is the photon transverse energy (a rectangular strip of $\Delta\eta\times\Delta\phi=0.015\times0.400$ is excluded from the track $\pt$ summation to allow for photons that convert into e$^+$e$^-$ pairs); (iii) the $\Sigma \et$ of ECAL energy surrounding the cluster within $0.06<\Delta R<0.40$ (and excluding a strip of $\Delta\eta\times\Delta\phi=0.04\times0.400$) must be less than $4.2\GeV + 0.006\et$; and (iv) the $\Sigma \et$ of HCAL energy surrounding the cluster within $0.15<\Delta R<0.40$ must be less than $2.2\GeV + 0.0025\et$. Here $\Delta R\equiv\sqrt{\Delta\phi^2+\Delta\eta^2}\equiv\sqrt{(\phi-\phi_\gamma)^2+(\eta-\eta_\gamma)^2}$.  We also require that the shower shape in $\eta$, $\sigma_{\eta\eta}$, be consistent with a photon.  The $\sigma_{\eta\eta}$ variable is a modified second moment of the electromagnetic energy cluster about its mean $\eta$ position, defined in Ref.~\cite{CMSQCDPhoton}. Topological and timing criteria suppress noise present in the ECAL~\cite{ECALPAS,ECALJINST}.

We reconstruct two photons using the selection described above and require that the invariant mass of the two photons satisfies $M_{\gamma\gamma}>60$~GeV.  The invariant mass and photon pseudorapidity selection criteria are optimized to produce the highest sensitivity for values of $M_{\rm S}$ and $n_{\rm ED}$ to which the present data is sensitive.  Alternating between fixing the $M_{\gamma\gamma}$ requirement and floating the $|\eta|$ requirement---and vice versa---results in a final choice of $|\eta|<1.44$ and $M_{\gamma\gamma}>500$~GeV.  The optimal choice for $|\eta|$ was very close to the ECAL barrel-endcap boundary, so we chose the edge of the transition region into the boundary for simplicity.  This defines the signal region.  The intervals $60<M_{\gamma\gamma}<200$~GeV and $200<M_{\gamma\gamma}<500$~GeV define the control and intermediate regions of the data, respectively.

The photon reconstruction and identification efficiency is measured in Monte Carlo (MC) simulation and corrected using a data/MC scale factor of $1.010 \pm 0.012$ derived from studying $Z\to e^+ e^-$ events.  The final efficiency is roughly constant as a function of the photon $\et$ and $\eta$.  The efficiency for an $\et>30$~GeV photon with $|\eta|<1.44$ is $(87.8\pm2.3)\%$, where the dominant systematic uncertainty is chosen to cover the variation as a function of $\et$ and $\eta$.  Therefore, the corresponding diphoton reconstruction and identification efficiency is $(77.1\pm4.5)\%$.

\section{Signal and Background Estimation}

We simulate ED in the ADD model using the \textsc{sherpa} (v1.1.2)~\cite{Sherpa} MC generator, which samples different operating points in $M_{\rm S}$ and $n_{\rm ED}$, followed by a fast parametric simulation of the CMS detector~\cite{FastSim}. A fast simulation is adequate for describing multiphoton final states and has been extensively validated using full simulation of the detector via {\sc geant4}~\cite{GEANT4}. The simulation includes both SM diphoton production and signal diphoton production via KK-graviton exchange, in order to account for the interference effects. We use {\sc cetq6l1}~\cite{CTEQ} parton distribution functions (PDF) in the simulation. The leading order (LO) \textsc{sherpa} cross sections are multiplied by a next-to-leading order (NLO) $K$ factor of $1.3 \pm 0.1$~\cite{EDNLO}, where the systematic uncertainty covers the variation of the $K$ factor with the diphoton mass.  Additionally, a $1.5\%$ relative uncertainty on the signal acceptance is included to account for uncertainty due to the PDF.

Backgrounds due to the mimicking of a photon signal by a jet are small in the signal region.  There are two sources of these backgrounds from isolated photon misidentification which we consider: multijet production and prompt photon production (i.e., photons from $\gamma+\text{jets}$).  In particular, we measure a misidentification rate, defined as the ratio of the number of isolated photons to non-isolated photons in a sample, where the non-isolated photons are selected similarly to the isolated photons except that they fail one of the isolation or shower-shape criteria.  The definition of the numerator and denominator are given so that the two sets are exclusive to each other.  The misidentification rate is measured in an EM enriched sample and is then applied to the observed events with one or more non-isolated photons, resulting in a prediction of the dijet and $\gamma+\text{jet}$ backgrounds.

Because a given control sample in which we measure the misidentification rate may contain some number of real, isolated photons that ``contaminate'' the misidentification-rate numerator, we correct for the numerator purity on a sample-by-sample basis.  This is done by releasing the $\sigma_{\eta\eta}$ requirement and fitting the numerator sample for the fraction of prompt photons using one-dimensional probability density histograms (``templates'') in $\sigma_{\eta\eta}$.  The signal template is constructed from MC simulation, and the background template is constructed by using inverted isolation criteria.  The measured misidentification rate falls from 28\% at $\et=30$ GeV to 2\% at $\et=120$~GeV.  We use two other complementary techniques (using converted photons and an isolation template to estimate prompt-photon contamination) to bound the misidentification rate and apply a conservative 20\% systematic uncertainty, which is the dominant uncertainty on the background estimation.

The diphoton background is computed with the \textsc{sherpa} MC program and then rescaled by an NLO $K$ factor of 1.3~\cite{ggNLO, EDNLO}.  This $K$ factor is alternatively derived with \textsc{diphox}~\cite{diphox}, wherein we observe that the $K$ factor decreases slowly as a function of diphoton invariant mass and stabilizes in the range of interest. We therefore use a $K$ factor of $1.3\pm0.3$ for the diphoton background to cover its variation as a function of $M_{\gamma\gamma}$ throughout the control, intermediate, and signal regions ($M_{\gamma\gamma}>60$~GeV). We observe (expect) 440 ($374 \pm 51$) events with $M_{\gamma\gamma}>60$~GeV and zero ($0.30 \pm 0.07$) with $M_{\gamma\gamma}>500$~GeV.

\begin{figure}[!Hhtb]
\centering
\includegraphics[angle=90,width=0.45\textwidth]{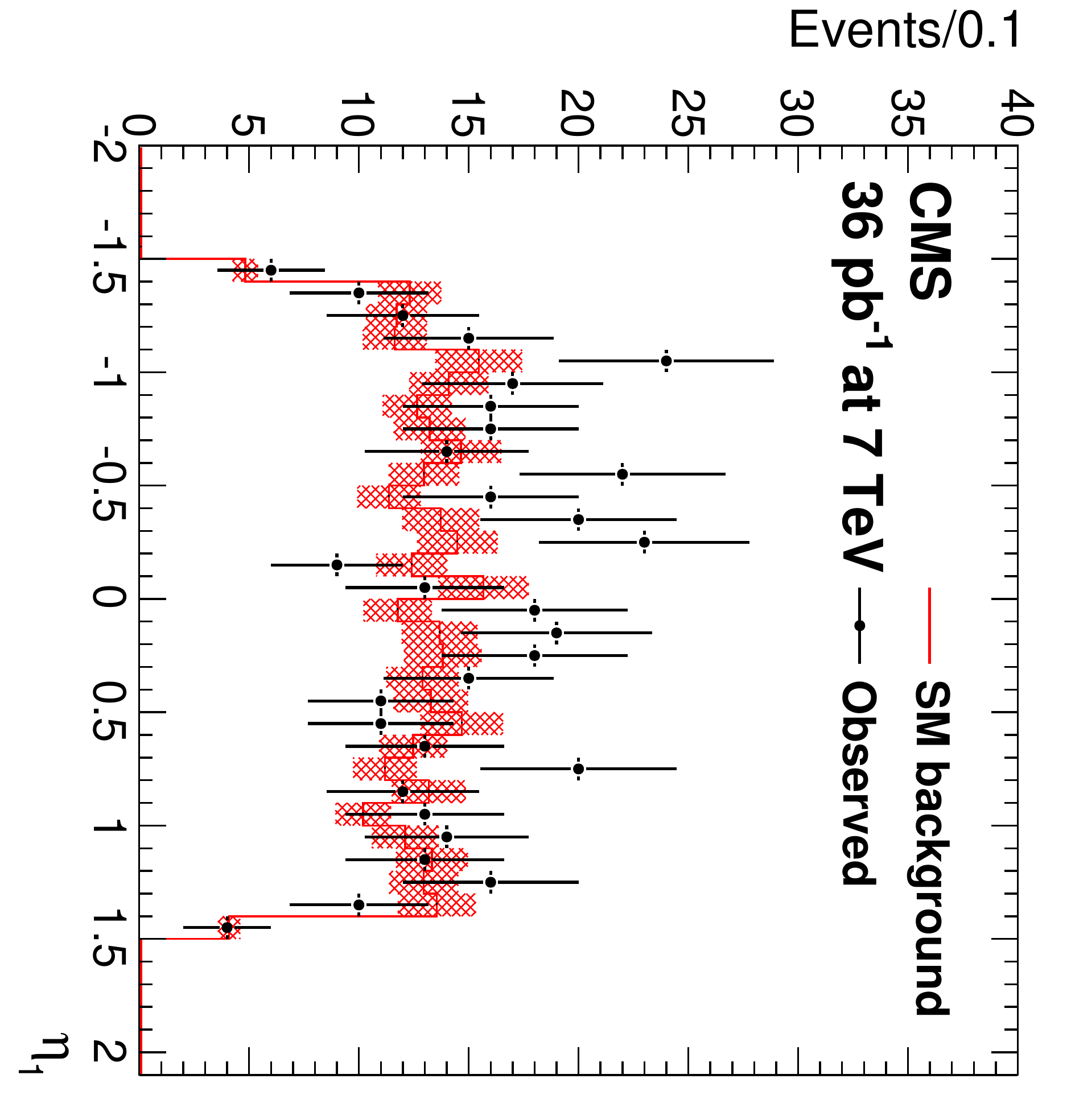}
\includegraphics[angle=90,width=0.45\textwidth]{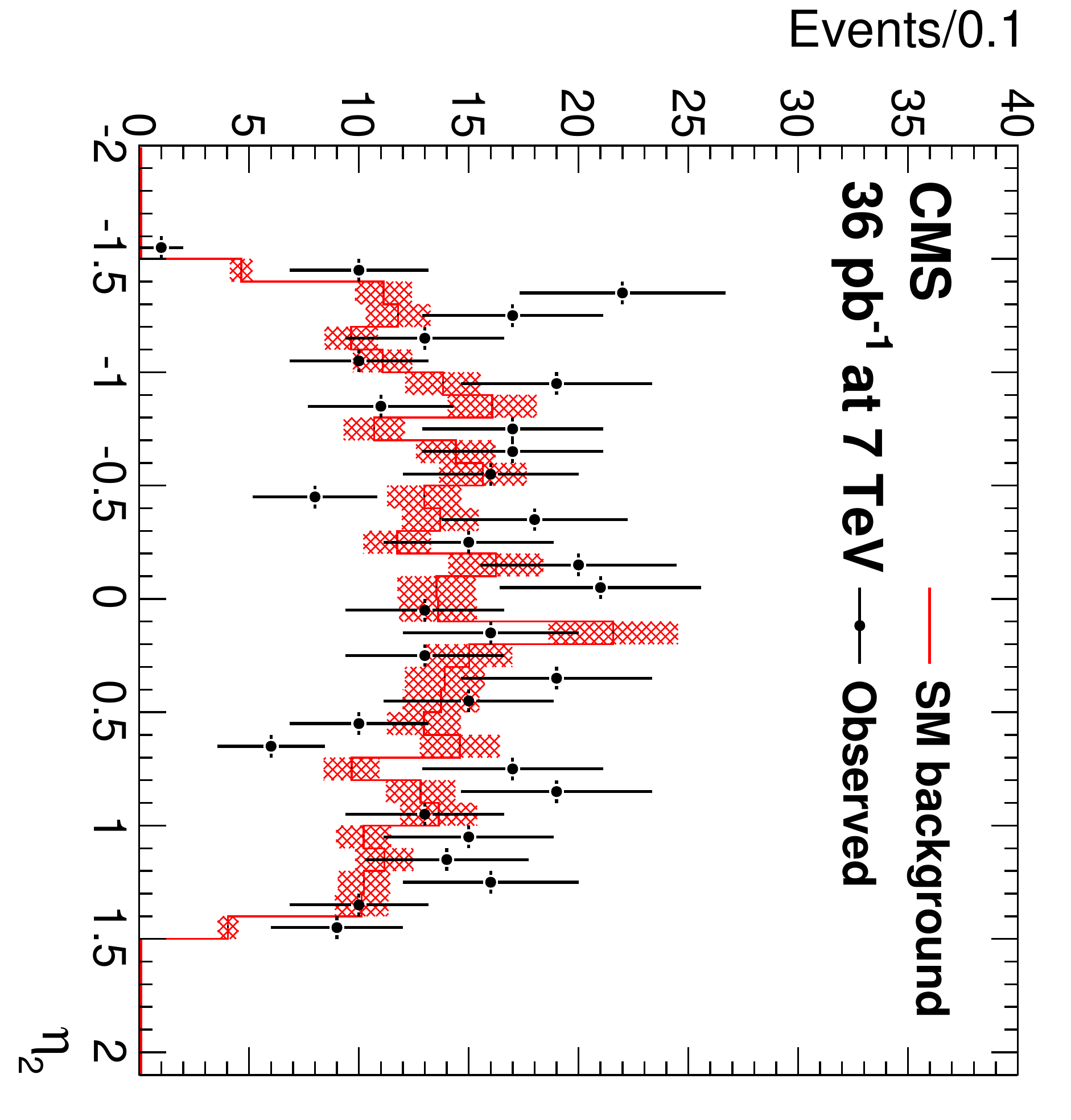}
\includegraphics[width=0.45\textwidth]{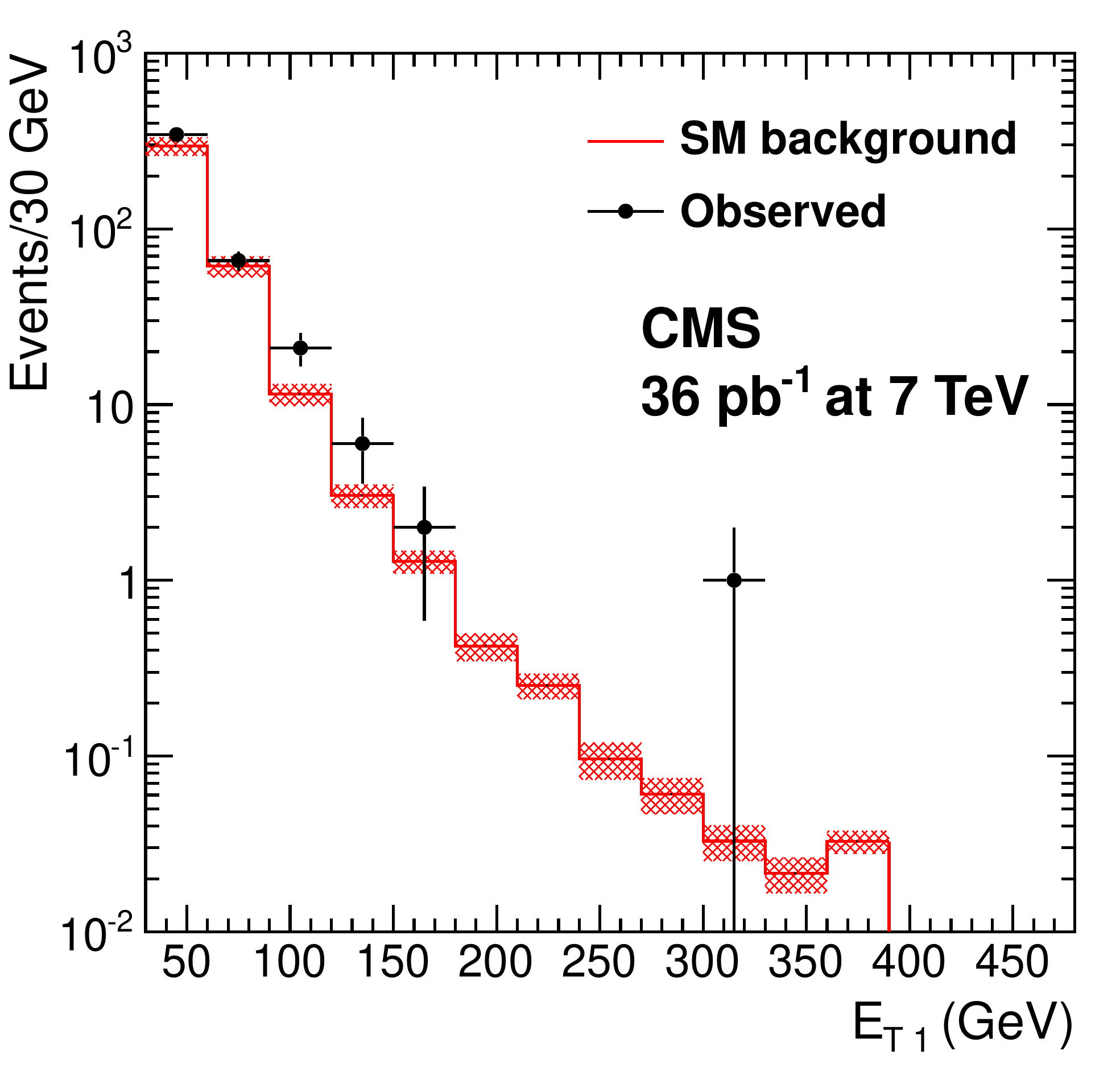}
\includegraphics[width=0.45\textwidth]{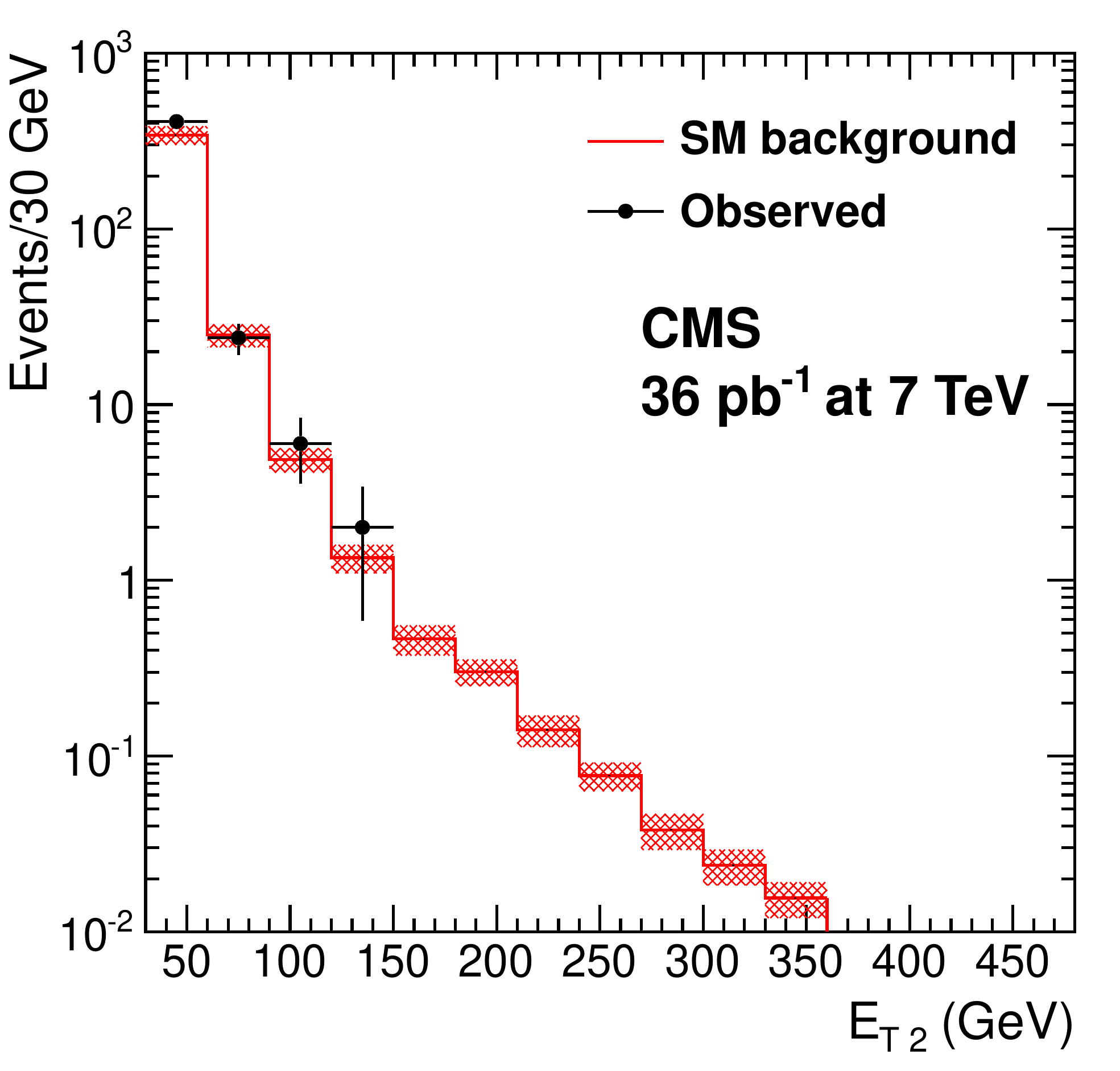}
\caption{Distributions in $\eta$ and $E_T$ for the leading and sub-leading photons. Points with error bars represent observed data; the solid histogram corresponds to the expected background from control regions.  Shading corresponds to the systematic uncertainty on the background expectation.}
\label{fig:etapt}
\end{figure}

\begin{figure}[!Hhtb]
\centering
\includegraphics[width=0.8\textwidth]{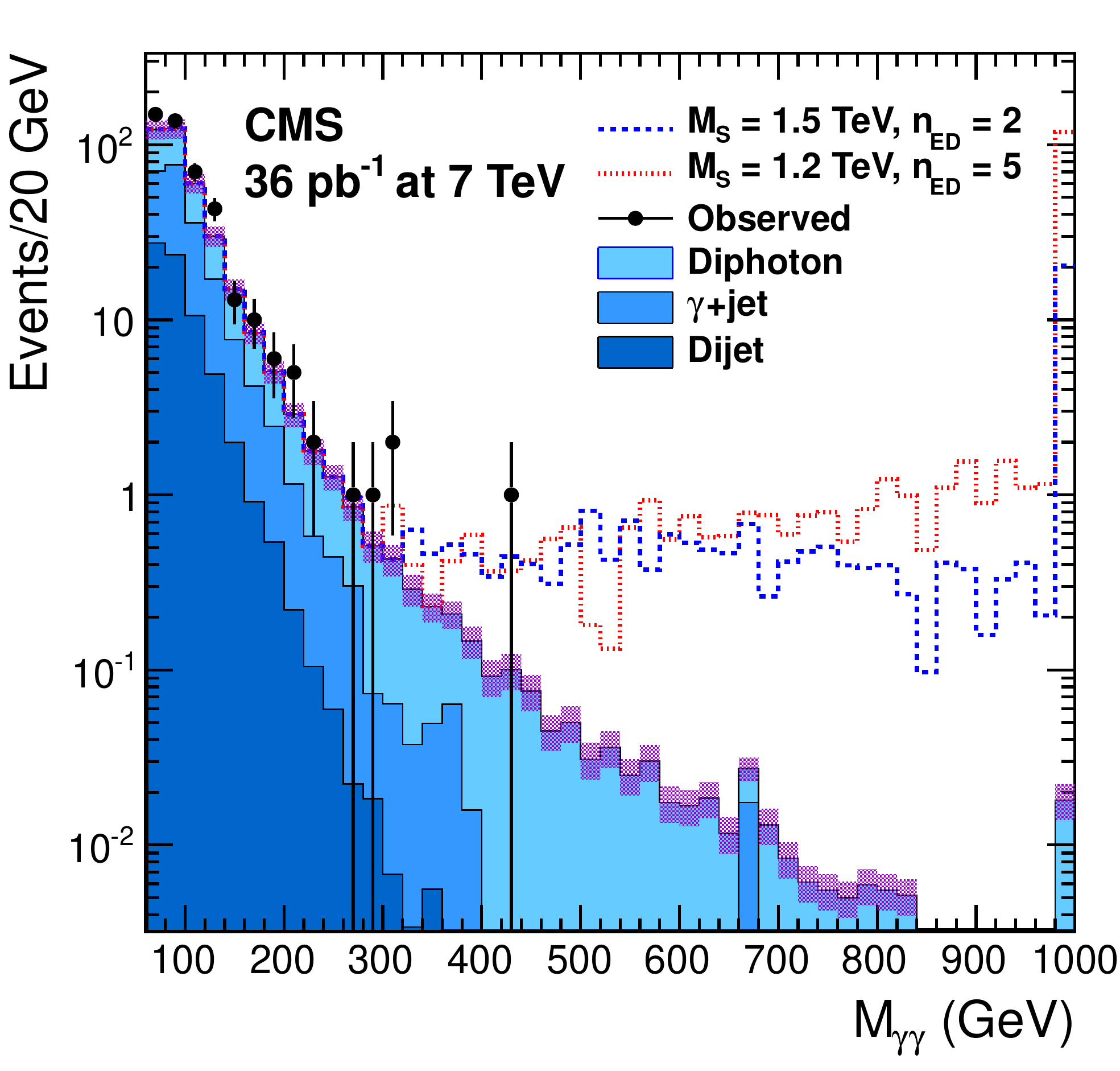}
\caption{Observed data (points with error bars) and background expectations (filled solid histograms) as a function of the diphoton invariant mass. Photons are required to be isolated, with $\et>30$~GeV and $|\eta|<1.44$.  Also shown with dashed lines are the signal distributions for two sets of model parameters.  Shaded bands around the background estimation correspond to systematic uncertainties.  The last bin is an overflow, including the sum of all contributions for $M_{\gamma\gamma}>1.0$~TeV.\label{fig:results_invmass}}
\end{figure}

Figure~\ref{fig:etapt} shows the pseudorapidity and transverse energy distributions for the photons with the largest and second-largest transverse energy in an event, along with the background predictions. Figure~\ref{fig:results_invmass} displays the invariant mass distribution for each of the backgrounds as well as the observed data, with the optimized $\eta$ requirement. Beyond $M_{\gamma\gamma}>200$~GeV, we see 12 events, consistent with $9.0\pm1.5$ expected background events.

Table~\ref{table:eventcontent} presents the number of observed data events and expected number of background events in the control, intermediate, and signal regions, respectively.  This table corresponds directly to the plot in Fig.~\ref{fig:results_invmass}.  In the control region, we find that the data are consistent with the background expectation within the systematic uncertainty.  The systematic uncertainty on the the total background takes into account the correlations between dijet and $\gamma+\text{jet}$ backgrounds arising because both depend on the same misidentification rate.  The relative combined uncertainty on the backgrounds in the signal region is 22\%, due nearly entirely to the diphoton NLO $K$-factor uncertainty.

\begin{table}[!h]
 \centering
\caption{Data yields and background expectations for reconstructed diphoton invariant mass ranges.  Full systematic uncertainties have been included.\label{table:eventcontent}} \smallskip
 \begin{tabular}{cccc}
 \hline
 \hline
 Process & $60<M_{\gamma\gamma}<200$~GeV & $200<M_{\gamma\gamma}<500$~GeV & $500$~GeV$<M_{\gamma\gamma}$ \\
\hline
 Dijets            & $70\pm28$  & $0.5\pm0.2$  & $0.0009\pm0.0004$ \\
 $\gamma+\text{Jets}$     & $145\pm7$  & $2.3\pm0.3$  & $0.016\pm0.003$ \\
 Diphotons         & $150\pm35$ & $6.2\pm1.4$  & $0.29\pm0.07$ \\
 Total Backgrounds & $365\pm49$ & $9.0\pm1.5$  & $0.30\pm0.07$ \\
 \hline
 Observed          & 428        & 12           & 0 \\
 \hline
 \hline
\end{tabular}
\end{table}

\begin{table}[!h]
\centering
\caption{Summary of systematic uncertainties.\label{table:systematics}}\smallskip
\begin{tabular}{ccc}
 \hline
\hline
  & Central Value & Relative Uncertainty \\
\hline
 Luminosity & 36 pb$^{-1}$ & 4.0\% \\
 Background diphoton $K$ factor & 0.30 Events & 23\% \\
 Signal Efficiency & 77.1\% & 6.0\% \\
 Signal diphoton $K$ factor & 1.3 & 7.7\% \\
\hline
 \hline
\end{tabular}
\end{table}

\section{Results}

We perform a counting experiment in the signal region ($M_{\gamma\gamma}>500$~GeV) and set 95\% CL upper limits on the quantity
\begin{equation}
S\equiv(\sigma_{\rm total}-\sigma_{\rm SM})\times\beta\times\mathcal{A},
\end{equation}
where $\sigma_{\rm total}$ represents the total diphoton production cross section (including both signal, SM, and interference effects), and $\sigma_{\rm SM}$ represents the SM diphoton production cross section, only.  We indicate the signal branching fraction to diphotons by $\beta$ and the signal acceptance by $\mathcal{A}$.  We utilize a standard Bayesian approach~\cite{PDG,D0Bayes} with a flat prior chosen for the signal cross section and log-normal priors for the nuisance parameters (integrated luminosity, signal efficiency, and background). The likelihood is constructed from the Poisson probability to observe $N$ events, given $S$, the signal efficiency of $(77.1\pm4.5)\%$, the expected number of background events ($0.30\pm0.07$), and the integrated luminosity $\mathcal{L}=(36\pm1)$~pb$^{-1}$~\cite{lumi}.  Table~\ref{table:systematics} summarizes the systematic uncertainties used as inputs to the limit calculation.

The observed (expected) 95\% CL upper limit on $S$ is 0.11 (0.13) pb. We translate the limit on $S$ into a limit on the parameters of the ADD model using the following technique. Because the effects of virtual-graviton exchange interfere with the SM diphoton production, generally, we expect the total diphoton cross section to have the following form:
\begin{equation}
        \sigma_{\rm total} = \sigma_{\rm SM} + \eta_G \; \sigma_{\rm int} + \eta_G^2 \; \sigma_{\rm ED},
        \label{eq:CS}
\end{equation}
where $\sigma_{\rm int}$ and $\sigma_{\rm ED}$ refer to the contributions to the total cross section from interference and direct ED effects, respectively.  Here, the dimensionful $\eta_G$ parameter specifies the strength of the ED effects and is related to $M_{\rm S}$ through      Eqs.~(\ref{eq:GRW}--\ref{eq:Hewett}).  Consequently, we parameterize $S$ as a function in the parameter $\eta_G$.  For the HLZ $n_{\rm ED} = 2$ case, $\eta_G$ depends on the signal spectrum due to an explicit $\hat{s}$ dependence in Eq.~(\ref{eq:HLZ}). Therefore, in this case we parameterize $S$ as a function of $1/M_{\rm S}^4$ and translate the limit on this parameter into a limit on $M_{\rm S}$.

\begin{figure}[!Hhtb]
\centering
\includegraphics[width=0.45\textwidth]{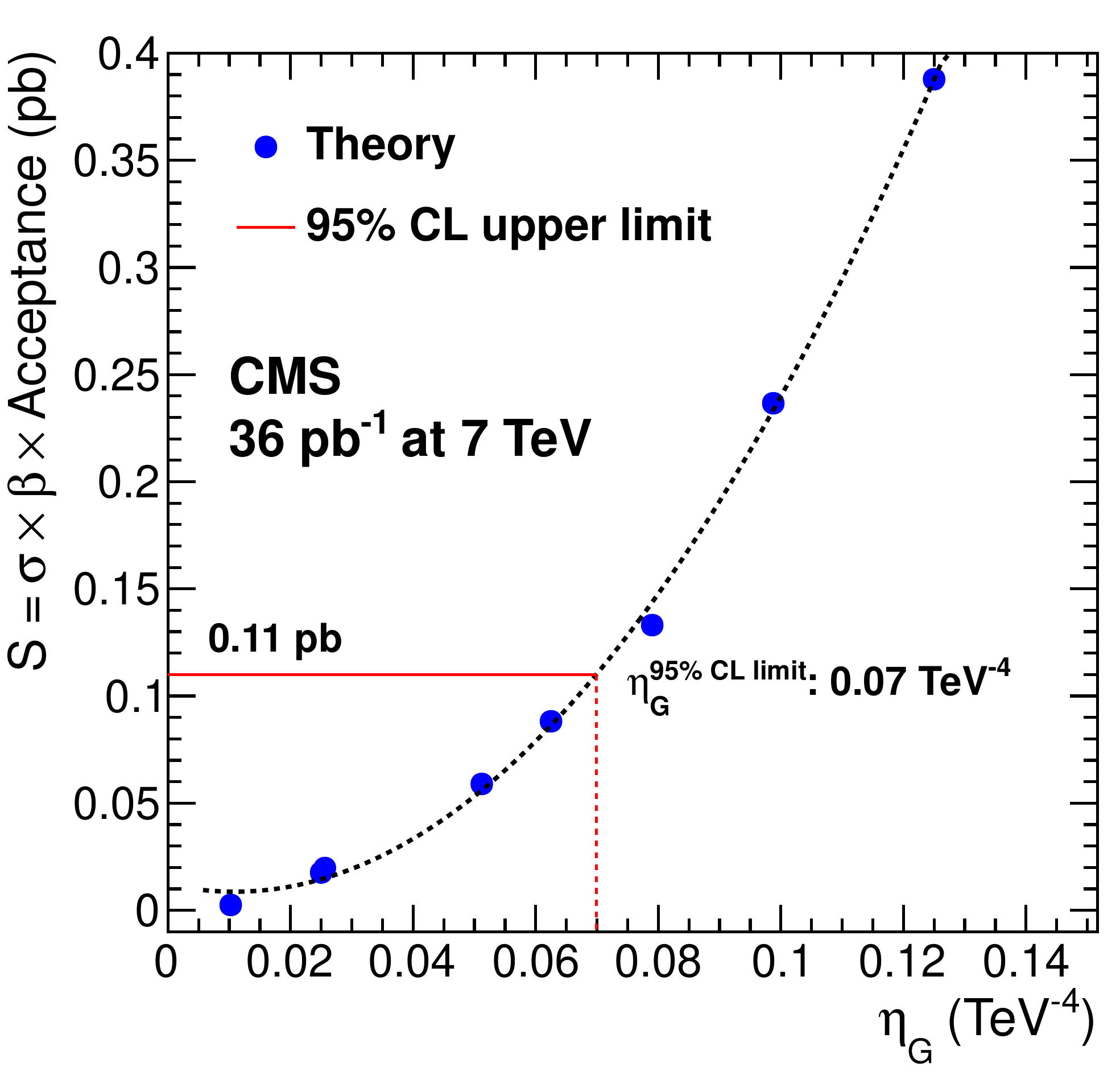}
\includegraphics[width=0.45\textwidth]{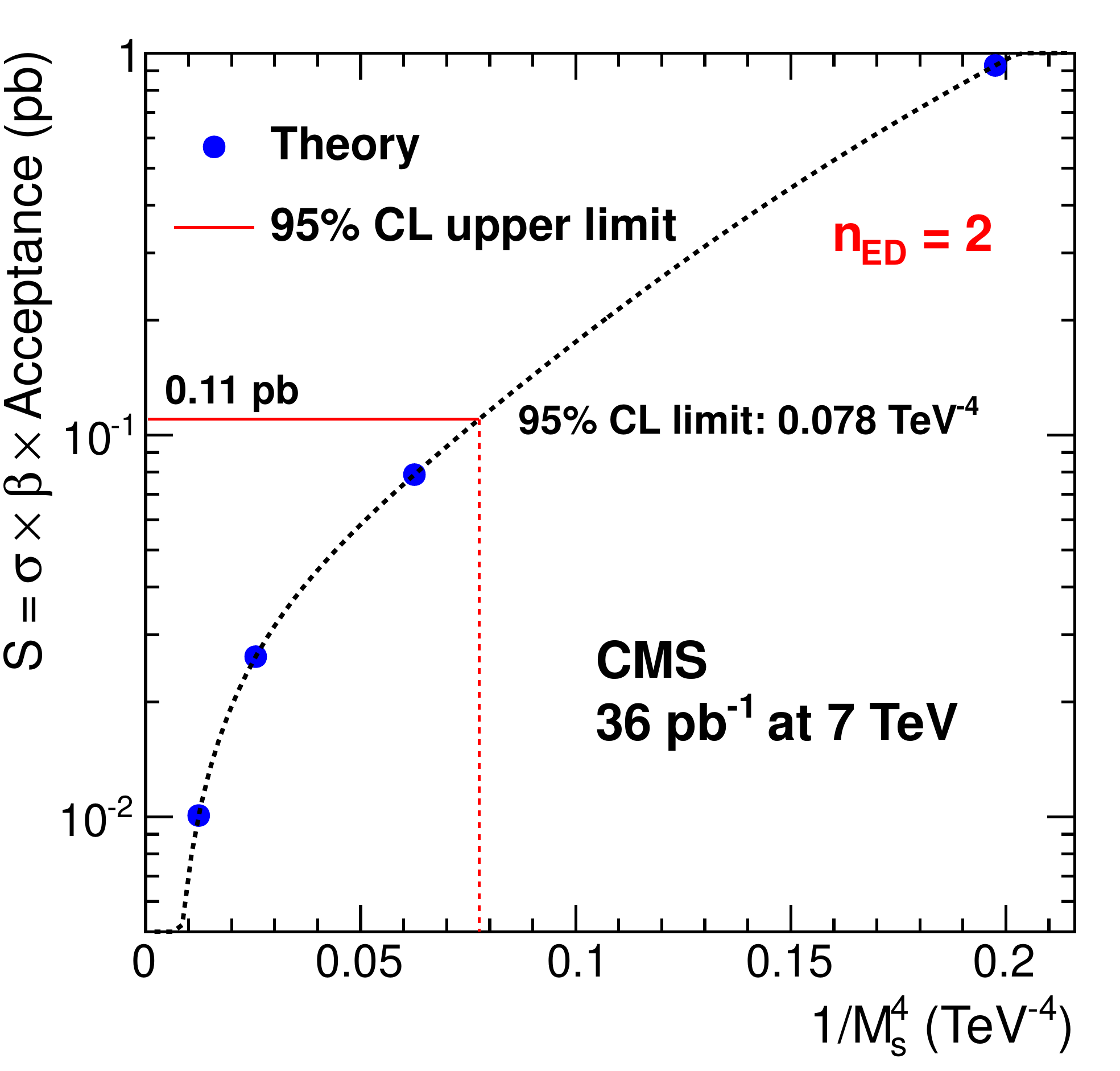}
\caption{Parameterization of $S$ as a function of the strength of the ED effect, $\eta_G$ for all cases except HLZ $n_{\rm ED}=2$ (left) and as a function of $1/M_{\rm S}^4$ for the $n_{\rm ED}=2$ case (right).  The parameterization shown is a fit according to Eq.~(\ref{eq:CS}).  The solid line is the 95\% CL exclusion limits on $S$, and is matched to the corresponding limits on $\eta_G$ and $1/M_{\rm S}^4$.\label{fig:cs-limit}}
\end{figure}

\begin{table}[!h]
\centering
\caption{$95\%$ CL limits on $M_{\rm S}$ (TeV), as a function of the convention and number of ED.  A comparison of the limits with a truncation of the production cross section above $\sqrt{\hat s}>M_{\rm S}$ is also shown.  The two limits for the Hewett convention correspond to positive and negative interference effects.\label{table:limits}}\smallskip
\begin{tabular}{c|c|cc|cccccc}
\hline
\hline
& GRW & \multicolumn{2}{c|}{Hewett} & \multicolumn{6}{c}{HLZ} \\
& & Pos. & Neg. & $n_{\rm ED}=2$ & $n_{\rm ED}=3$ & $n_{\rm ED}=4$ & $n_{\rm ED}=5$ & $n_{\rm ED}=6$ & $n_{\rm ED}=7$ \\
Full  & 1.94 & 1.74 & 1.71 & 1.89 & 2.31 & 1.94 & 1.76 & 1.63 & 1.55 \\
Trunc. & 1.84 & 1.60 & 1.50 & 1.80 & 2.23 & 1.84 & 1.63 & 1.46 & 1.31 \\
\hline
\hline
\end{tabular}
\end{table}

The observed 95\% CL upper limit on $S$ together with the parameterization of $S$ as a function of $\eta_G$ and $1/M_{\rm S}^4$ are shown in the left pane of Fig.~\ref{fig:cs-limit}. The intersection of the limit on $S$ with the curves determines the upper 95\% CL upper limits on the parameters $\eta_G$ and $1/M_{\rm S}^4$. As seen from these plots, the upper limits are equal to $0.070$~TeV$^{-4}$ for $\eta_G$ and $0.078$~TeV$^{-4}$ for $1/M_{\rm S}^4$ (for $n_{\rm ED}=2$). We further translate these limits, by means of Eq.~(\ref{eq:HLZ}), into lower bounds on the fundamental Planck scale for various numbers of extra dimensions, $n_{\rm ED}$, as shown in Table~\ref{table:limits}. This is calculated trivially for $n_{\rm ED} = 2$ and for $n_{\rm ED}>2$ by using Eq.~(\ref{eq:HLZ}). The limits in convention~\cite{GRW} are identical to the HLZ limits for $n_{\rm ED}=4$; the limit for the Hewett convention with constructive interference is 1.74~TeV and is close to the HLZ limit for $n_{\rm ED} = 5$.

We note that the LO signal cross section calculations become non-perturbative when the value of $\hat s$ in the $2 \to 2$ process exceeds $M_{\rm S}^2$. This effect is not taken into account in the \textsc{sherpa} cross section calculations used in this analysis, or in previous studies of this process at the Tevatron~\cite{D0diphoton}, where the effect is not expected to be important due to the lower collider energy. Because the energy of the LHC is significantly higher than the limits on $M_{\rm S}$ we are able to set in this analysis, we take this effect into account by conservatively assuming that the signal cross section is zero for $\sqrt{\hat s} > M_{\rm S}$. Under these assumptions, the limits on $M_{\rm S}$ decrease by 5\% for $n_{\rm ED}=2$ (1.80 TeV)  and 15\% for $n_{\rm ED}=7$ (1.31 TeV).  A summary of the limits under the assumption of a truncated production cross section is also shown in Table~\ref{table:limits}.

In addition to setting limits on a specific model of large extra dimensions, we can also set a model-independent limit on any new physics model which results in central, high-$\et$ diphotons, either resonant or non-resonant (e.g., Kaluza--Klein gravitons in the Randall--Sundrum model \cite{RS}).  We measure a 95\% CL exclusion on the cross section times branching fraction times acceptance of 0.110 pb, for diphoton pairs with $M_{\gamma\gamma}>500$~GeV and the following kinematic requirements on each of the two photons: $\et > 30$~GeV and $|\eta|<1.44$.

\section{Conclusions}

In conclusion, we have performed a search for large extra dimensions~\cite{add,add1} in the diphoton final state with a data sample collected in pp collisions at $\sqrt{s} = 7$ TeV corresponding to an integrated luminosity of 36 pb$^{-1}$.  We optimize the signal selection to reach maximum sensitivity in a counting experiment in a one-sided mass window by selecting events with centrally produced photons and large diphoton invariant mass.  Given the absence of an excess over the SM direct diphoton background, we set lower limits on the cutoff scale $M_{\rm S}$ in the range 1.6--2.3 TeV.  While this analysis was being finalized, a phenomenological interpretation of the dijet angular distribution results from the  CMS and ATLAS experiments appeared \cite{dijetpheno} and suggested even stronger limits on $M_{\rm S}$.  However, a dedicated experimental analysis and interpretation of the dijet data in the models with large extra dimensions has yet to be conducted.  The results presented in this Letter extend the current limits reached at the Tevatron~\cite{D0diphoton,D0dijet} in all but the $n_{\rm ED}=2$ case.

\section*{Acknowledgments}
We wish to congratulate our colleagues in the CERN accelerator departments for the excellent performance of the LHC machine. We thank the technical and administrative staff at CERN and other CMS institutes, and acknowledge support from: FMSR (Austria); FNRS and FWO (Belgium); CNPq, CAPES, FAPERJ, and FAPESP (Brazil); MES (Bulgaria); CERN; CAS, MoST, and NSFC (China); COLCIENCIAS (Colombia); MSES (Croatia); RPF (Cyprus); Academy of Sciences and NICPB (Estonia); Academy of Finland, ME, and HIP (Finland); CEA and CNRS/IN2P3 (France);  BMBF, DFG, and HGF (Germany); GSRT (Greece); OTKA and NKTH (Hungary); DAE and DST (India); IPM (Iran); SFI (Ireland); INFN (Italy); NRF and WCU (Korea); LAS (Lithuania); CINVESTAV, CONACYT, SEP, and UASLP-FAI (Mexico); PAEC (Pakistan); SCSR (Poland); FCT (Portugal); JINR (Armenia, Belarus, Georgia, Ukraine, Uzbekistan); MST and MAE (Russia); MSTD (Serbia); MICINN and CPAN (Spain); Swiss Funding Agencies (Switzerland); NSC (Taipei); TUBITAK and TAEK (Turkey); STFC (United Kingdom); DOE and NSF (USA).

\bibliography{auto_generated}   
\cleardoublepage\appendix\section{The CMS Collaboration \label{app:collab}}\begin{sloppypar}\hyphenpenalty=5000\widowpenalty=500\clubpenalty=5000\textbf{Yerevan Physics Institute,  Yerevan,  Armenia}\\*[0pt]
S.~Chatrchyan, V.~Khachatryan, A.M.~Sirunyan, A.~Tumasyan
\vskip\cmsinstskip
\textbf{Institut f\"{u}r Hochenergiephysik der OeAW,  Wien,  Austria}\\*[0pt]
W.~Adam, T.~Bergauer, M.~Dragicevic, J.~Er\"{o}, C.~Fabjan, M.~Friedl, R.~Fr\"{u}hwirth, V.M.~Ghete, J.~Hammer\cmsAuthorMark{1}, S.~H\"{a}nsel, M.~Hoch, N.~H\"{o}rmann, J.~Hrubec, M.~Jeitler, G.~Kasieczka, W.~Kiesenhofer, M.~Krammer, D.~Liko, I.~Mikulec, M.~Pernicka, H.~Rohringer, R.~Sch\"{o}fbeck, J.~Strauss, F.~Teischinger, P.~Wagner, W.~Waltenberger, G.~Walzel, E.~Widl, C.-E.~Wulz
% ...
\vskip\cmsinstskip
% ...
1:~~Also at CERN, European Organization for Nuclear Research, Geneva, Switzerland\\
% ...

\end{sloppypar}
\end{document}